\documentclass[authoryear]{cas-sc}

\usepackage[authoryear]{natbib}

\usepackage{graphicx}
\usepackage{amssymb}
\usepackage{longtable}
\usepackage{multirow}
\usepackage{appendix}
\usepackage[T1]{fontenc}
\usepackage{lscape}

\usepackage{amsmath}
\usepackage{caption} 
\usepackage{subcaption} 
\usepackage{booktabs}
\usepackage{tabularx}
\usepackage{algorithm2e}
\usepackage{vcell}
\usepackage{hhline}
\usepackage{balance}
\usepackage{float}
\usepackage{hyperref}
\usepackage{url}
\usepackage{ulem}
\usepackage{graphicx}
\usepackage{xcolor} 

\def\tsc#1{\csdef{#1}{\textsc{\lowercase{#1}}\xspace}}
\tsc{WGM}
\tsc{QE}
\tsc{EP}
\tsc{PMS}
\tsc{BEC}
\tsc{DE}

\begin{document}
\let\WriteBookmarks\relax
\def\floatpagepagefraction{1}
\def\textpagefraction{.001}

\shorttitle{}

\shortauthors{}


\title [mode = title]{Prediction of Retention Time in Larger Antisense Oligonucleotide Datasets using Machine Learning} 



%
\author[1]{Manal Rahal}



\ead{manal.rahal@kau.se}


\affiliation[1]{organization={Department of Mathematics and Computer Science},
    addressline={}, 
    city={Karlstad},
    citysep={}, 
    postcode={651 88}, 
    state={},
    country={Sweden}}

\author[1]{Bestoun S. Ahmed}
\ead{bestoun@kau.se}

\author[2]{Christoph A. Bauer}[orcid =0000-0002-6035-6944]
\ead{christoph.bauer@astrazeneca.com}

\author[2]{Johan Ulander}
\ead{johan.ulander@astrazeneca.com}

\affiliation[2]{organization={Data Science and Modelling, Pharmaceutical Sciences, R\&D, AstraZeneca},
    addressline={}, 
    city={Gothenburg},
    state={},
    country={Sweden}}

\author[3]{Jörgen Samuelsson}
\ead{jorgen.samuelsson@kau.se}

\affiliation[3]{organization={Department of Engineering and Chemical Sciences},
    addressline={}, 
    city={Karlstad},
    citysep={}, 
    postcode={651 88}, 
    state={},
    country={Sweden}}





\begin{abstract}
 Antisense oligonucleotides (ASOs) are nucleic acid molecules with transformative therapeutic potential, especially for diseases that are untreatable by traditional drugs. However, the production and purification of ASOs remain challenging due to the presence of unwanted impurities. One tool successfully used to separate an ASO compound from the impurities is ion pair liquid chromatography (IPC). It is a critical step in separation, where each compound is identified by its retention time ($t_\mathrm{R}$) in the IPC. Due to the complex sequence-dependent behavior of ASOs and variability in chromatographic conditions, the accurate prediction of $t_\mathrm{R}$ is a difficult task. This study addresses this challenge by applying machine learning (ML) to predict $t_\mathrm{R}$ based on the sequence characteristics of ASOs. Four ML models—Gradient Boosting, Random Forest, Decision Tree, and Support Vector Regression- were evaluated on three large ASOs datasets with different gradient times. Through feature engineering and grid search optimization, key predictors were identified and compared for model accuracy using root mean square error, coefficient of determination R-squared, and run time. The results showed that Gradient Boost performance competes with the Support Vector Machine in two of the three datasets, but is 3.94 times faster to tune. Additionally,  newly proposed features representing the sulfur count and the nucleotides residing at the first and last positions of a sequence found to improve the predictive power of the models. This study demonstrates the advantages of ML-based $t_\mathrm{R}$ prediction at scale and provides insights into interpretable and efficient utilization of ML in chromatographic applications.  
\end{abstract}


\maketitle

\section{Introduction}\label{sec:introduction}

Oligonucleotides are nucleic acid-based molecules that target undruggable proteins in the human body for therapeutic purposes \cite{Thakur2022}. Among them, Antisense oligonucleotides (ASOs), which are short, chemically modified, single-stranded DNA or RNA-based molecules that can target specific genes in the human system and influence their activity (\cite{ENMARK2022}, \cite{ROBERTS2020}). Recent advances in molecular biology have been mainly associated with the use of ASOs in advanced therapeutic applications \cite{Egli2023}. Despite their transformative therapeutic potential, ASO production remains complex due to the presence of large amounts of impurities. These impurities must be analyzed and removed before an ASO can be administered as a drug, which significantly increases analysis costs. Given the innovative benefits of ASOs and the complexity of production, several approaches have been explored to optimize the analysis process. Among these approaches, machine learning (ML) is one of the most promising methods in terms of prediction accuracy.

\begin{figure*}
    \centering
         \includegraphics[scale=.45]{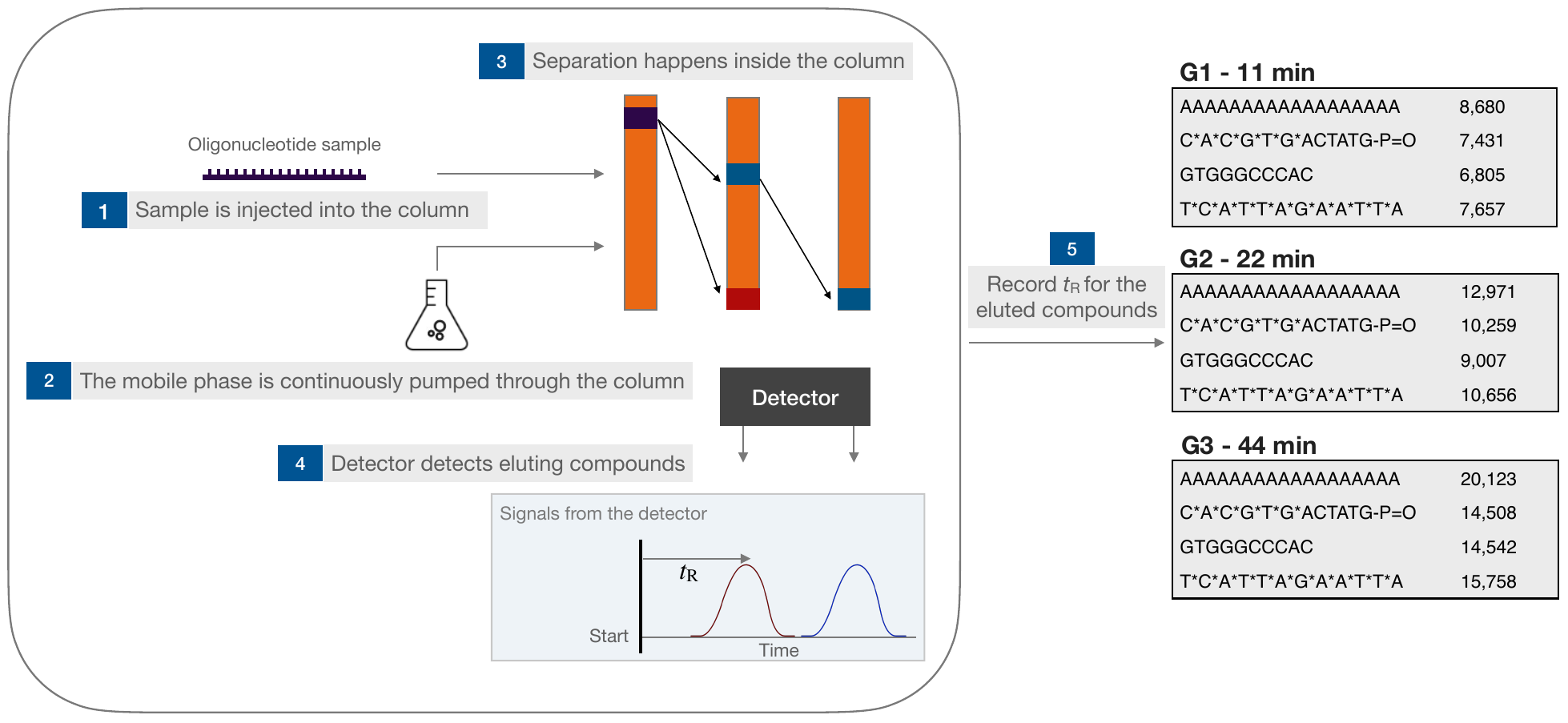}
	\caption{Demonstration of the separation process of an ASO sample through IPC. Chemical modification is denoted by *.}
	\label{fig:Chromatography}
\end{figure*}

The separation process is considered difficult to perform analytically, given the nature of ASO and its closely related impurities \cite{ENMARK2022}. One of the most commonly used techniques to identify an ASO full-length product (FLP) and the produced impurities is ion-pair liquid chromatography (IPC) coupled with mass spectrometry \cite{ENMARK2022}. In IPC, the time a compound spends in the column before eluting is called retention time ($t_\mathrm{R}$), it is calculated from injection time to elution \cite{Moruz2017}. The method involves distributing the compounds between two phases, a mobile phase containing a mixture of water and an organic solvent that is continuously pumped through the stationary phase. Figure \ref{fig:Chromatography} illustrates the separation and identification steps in an IPC system.

First, the ASO sample is injected into the column, where the mobile phase helps to carry the sample through. As they propagate through the column, each compound in the mixture interacts differently with the stationary phase, resulting in separation. Some compounds elute early, while others strongly interact with the stationary phase and are retained longer inside the column. In this case, the amount of organic solvent is increased to decrease the strong interaction and control the $t_\mathrm{R}$. The change in the amount of the organic solvent over time is called a gradient. As the compounds exit the system, they are detected. The recorded $t_\mathrm{R}$ depends on sequence-specific properties and is measured at the peak of the signal. The uniqueness property of the $t_\mathrm{R}$ allows for the identification of compounds.

Recent studies have explored empirical and ML methods for predicting $t_\mathrm{R}$. This is feasible given the correlation between $t_\mathrm{R}$ and nucleotide bases that form an ASO sequence. Mathematical models rely on nucleotide composition but struggle under varying experimental conditions. On the other hand, ML models, especially Support Vector Regression (SVR), have demonstrated improved performance due to their ability to model non-linear relationships. In an advanced stage, predicting $t_\mathrm{R}$ would inform the selection of the optimal conditions for the separation of oligonucleotide compounds \cite{ENMARK2022}. As a result of knowing the optimal conditions for analysis, a controlled and reduced number of physical experiments are then required \cite{SAMUELSSON2019}, thus significantly reducing experimentation costs and accelerating research in this field.

The ability to transform the physiochemical properties of an ASO sequence into numerical quantities that an ML model understands offers remarkable prospects for optimization. Although current prediction approaches provide satisfactory results, there is a lack of understanding of their predictive power compared to other classically used ML models. Additionally, these models are often trained on small datasets and lack comparative evaluations with alternative ML approaches on larger ASO datasets. One of the key focuses of this paper is to address this gap and evaluate the performance of various ML models compared to the methods currently in use. The second point of interest is to understand the influence of multiple features on the prediction of $t_\mathrm{R}$ and to identify the most important predictors.

This paper aims to show the application of ML for predicting $t_\mathrm{R}$ in large datasets of synthetic ASO compounds. Given that the current application of ML in the case of $t_\mathrm{R}$ prediction is limited to specific models, this paper presents a comparative perspective to fill this gap. It also aims to analyze and compare the performance of the various ML models in terms of multiple evaluation criteria. ML models are applied to three synthetic ASO datasets, where performance is evaluated according to three evaluation criteria: root mean square error (RMSE), R-squared coefficient of determination ($R^2$), and run time. The input data are identically pre-processed before the application of each model to ensure a fair comparison. 

The contribution of this paper is summarized as follows:
\begin{enumerate}
  \item {Provide a comprehensive view on the use of ML in the optimization of chromatography applications}
  \item Perform in-depth analysis of the performance of multiple ML models to predict $t_\mathrm{R}$ in large ASO datasets.
  \item Evaluate the impact of newly proposed features on the prediction of $t_\mathrm{R}$. 
\end{enumerate}

The remainder of this paper is organized as follows. Section \ref{sec:relatedwork} describes the work done in the application of $t_\mathrm{R}$ prediction, highlighting the common use of the support vector machine (SVM) method in the literature. Since this paper follows an experimental approach, the experimental dataset is described in Section \ref{sec:dataset} and the work methodology is described in Section \ref{sec:method}. Section \ref{sec:results} presents the results of the performance-based evaluation and the analytically deduced conclusions. Section \ref{sec:conclusion} presents the overall summary of the work.

\section{Related work}\label{sec:relatedwork}

The prediction of $t_\mathrm{R}$ in IPC has various applications besides protein identification. In particular, predictions are used to optimize experiments in \textit{silico} before performing physical experiments. In this context, digital simulations are critical to narrow down the number of physical experiments performed. Combined with the prediction of other chromatography parameters, chemists can easily learn useful information about the behavior of the compounds. Therefore, partial or full automation of this process is much needed to improve the ASOs analysis methods and infer valuable insights. For these reasons, many researchers have investigated the prediction of the retention behavior of nucleotide sequences. The research efforts investigated two types of nucleic acid compounds (peptides and oligonucleotides) using a wide range of methods (mathematical and ML approaches) and various composition modifications (such as phosphorothioated and non-phosphorothioated). 

To date, the different approaches to predict the $t_\mathrm{R}$ of oligonucleotides are few compared to peptides \cite{ENMARK2022}. In the literature, there are only a few notable models that predict $t_\mathrm{R}$ for IPC separation of oligonucleotides, including \cite{GILAR2002}, \cite{Strum2007}, \cite{studzinska2015}, \cite{LIANG2018}, and \cite{Kohlbacher2006}. Researchers used multiple approaches that can be grouped into non-ML and ML-based methods. In fact, the oligonucleotide prediction methods started and evolved with simple mathematical models, most notably the model proposed in \cite{GILAR2002}. The mathematical model considers five input variables, mainly representing the amount of each nucleotide and the total length of the sequence. The model then calculates the contribution of each nucleotide and predicts the $t_\mathrm{R}$. Gilar's model performed well at high temperatures. However, the accuracy decreased significantly at low temperatures because of the influence of low temperatures on the compound's structure, making it difficult to predict. Despite their advantages of ease of use and effectiveness with small datasets, non-ML methods did not deliver excellent prediction performance \cite{ENMARK2022}. In particular, these models have limitations in fully leveraging the potential of all features within the oligonucleotide sequence.

Nevertheless, Gilar \textit{et al.}'s successful mathematical model \cite{GILAR2002} was considered a notable milestone in advancing research in this field. The limitations in \cite{GILAR2002} motivated other researchers, such as in \cite{Kohlbacher2006, Strum2007}, to develop more generalized models based on machine learning, inspired by the use of Support Vector Machines (SVM) in peptides. Although the SVM model was trained on only 72 oligonucleotide sequences, it has significantly improved prediction accuracy compared to the non-ML approaches. This improvement was due to the model’s capability to represent nonlinear relationships \cite{Strum2007}. In fact, based on the successful application of SVM in \cite{Kohlbacher2006}, SVM was later used extensively in $t_\mathrm{R}$ prediction studies, typically using small, curated datasets.

In this paper, ML-based methods in the literature have been examined and their performance and applications evaluated to provide a comparative perspective for our approach. ML-based approaches were first used to predict the $t_\mathrm{R}$ for peptides, which are molecular compounds that share common characteristics with oligonucleotides. One of the earliest ML-based studies is \cite{Petritis2003}, where a neural network was trained on approximately 7000 peptides to predict the elution time of 5200 peptides using 20 input features. However, the availability of large experimental data to train the NN significantly limited the practicality of applying ML \cite{Moruz2010}. Therefore, the subsequent work in (\cite{Klammer2007}, \cite{Moruz2010}) proposed variations of the SVM-based models that can be adequately trained using a small dataset of peptides. In \cite{Klammer2007}, the SVR model was introduced as a generalized adaptable approach that predicts with high accuracy, regardless of the experimental conditions. Most importantly, it could achieve high-accuracy predictions using a small training dataset. In \cite{Moruz2010}, a new approach to the prediction of $t_\mathrm{R}$ was presented. Unlike the approach to use experimentally pre-identified peptides to train an ML model, a two-step selection and calibration approach was proposed in \cite{Moruz2010}. For the selection stage, a sample denoted as a control sample, consisting of 50 to 100 peptides, was separated under specific experimental conditions. The resulting observed retention times are then used to select a model from a collection of pre-trained SVR models. The control sample was also used to tune the selected model for the experimental conditions that are being evaluated. Once the ML model has been selected and calibrated, it was used to predict the $t_\mathrm{R}$ on unseen data. Consequently, these advances in prediction approaches for peptides fundamentally guided research efforts to predict the $t_\mathrm{R}$ of oligonucleotides at a later stage. 

Following the successful use of SVR in peptides and non-phosphorothioated oligonucleotides, the model was used to predict the $t_\mathrm{R}$ for phosphorothioated ASOs in \cite{ENMARK2022}. The dataset was around 100 ASO compounds, where the influence of three different gradient runs on the retention behavior was investigated. It is important to note that phosphorothioated ASOs are chemically modified oligonucleotides in which the oxygen atom is replaced with a sulfur atom in the backbone of the phosphate group, resulting in a more stable compound \cite{Khvorova2017}. In contrast to \cite{Strum2007}, the frequencies of nucleotide bases A, T, C, and G in a sequence were found to be more important than other extracted features. In addition, when compared with Gilar \textit{et al.} model, SVR remarkably performed better at accurately predicting $t_\mathrm{R}$, achieving lower RMSE at all gradient slopes.

Recent advancements in ML approaches have further improved $t_\mathrm{R}$ prediction. Liu \textit{et al.} \cite{Liu2024} leveraged deep learning models with large-scale data sets such as MassBank to improve the prediction accuracy of small molecule retention in liquid chromatography, showcasing the power of deep learning to manage variability and standardization challenges. Randazzo \textit{et al.} applied ML models, specifically XGBoost, to predict $t_\mathrm{R}$ for small molecules in nano-HPLC, highlighting that ensemble approaches significantly improved prediction accuracy compared to traditional methods \cite{Randazzo2020}. Furthermore, Wolfer \textit{et al.} showed that the use of ML in UPLC-MS for the identification of metabolites in untargeted profiling was highly effective, further supporting the importance of model choice and optimization for chromatographic analysis \cite{Wolfer2016}.

To date, most research in $t_\mathrm{R}$ prediction for ASOs has focused on small datasets (typically around 100 compounds) and relied primarily on SVR models. While SVR is efficient and has demonstrated high accuracy, little is known about how it compares with other ML models in larger ASO datasets under varying experimental conditions.

This paper aims to fill this gap by evaluating and comparing four ML models (SVR, Gradient Boosting, Random Forest, and Decision Tree) on three large ASO datasets. It also investigates the impact of newly proposed sequence-derived features on the prediction accuracy.

\section{Methodology}\label{sec:method}

This section outlines the dataset and the ML pipeline used to predict the $t_\mathrm{R}$ for the ASO compounds described in Section \ref{sec:dataset}. The rationale for selecting specific predictive features, the motivation behind the choice of ML models, and the overall experimental design is thoroughly described. All steps, including data preprocessing, feature extraction, model training, hyperparameter tuning, and performance evaluation, were implemented in Python 3.9.7 using Scikit-learn 0.24.2 and other relevant libraries.

\subsection{Dataset}\label{sec:dataset}

The data used in this experiment are obtained from three chromatography experiments on three different gradient times, gradient one (G1 = 11 min), gradient two (G2 = 22 min), and gradient three (G3 = 44 min). In analytical chemistry, the selection of the gradient time for separation is determined by balancing the efficiency of separation and cost. The separation scientist aims to extrapolate the optimal condition by finding a sufficiently good separation performance within a short time. A narrower range would make the construction of three distinct gradient models more challenging, and a wider range would affect the execution time of the experiments. Therefore, G2 was selected as a balanced reference for both run time and separation performance. Then, a shorter (G1) and a longer time (G3) were selected to ensure enough variability in the chromatography conditions, leading to sufficient differences in the performance of the different ML models for each gradient time. However, a very short time will not be good enough from a chemistry point of view, as separation will compounds will not separate as they should. The change in the gradient concentration results in a new $t_\mathrm{R}$ for each unique compound, as illustrated in Figure \ref{fig:Chromatography}. The $t_\mathrm{R}$ of the compounds, recorded at the three gradient runs, is the target variable in this experiment.

During each of the gradients, 45 different FLPs of ASOs compounds go through the separation process, where each FLP loses one nucleotide at a time, and a new compound is produced. As the newly separated compound, commonly known as a shortmer, exits the system, it is detected by the detector as a signal. Then $t_\mathrm{R}$ is calculated for every detected signal, as illustrated in Figure \ref{fig:Chromatography}. This process results in the collection of a total of 876 non-null $t_\mathrm{R}$ data points corresponding to FLPs and their separated shortmers. The recorded $t_\mathrm{R}$ is specific to the chosen gradient run, thus resulting in three different datasets of ASO compounds and $t_\mathrm{R}$ pairs, one for each gradient run. As a result, the size of the data in G1 and G2 is 876 data points, while G3 includes 870 non-null data points. An example of an FLP and its separated compounds is illustrated in Table \ref{Sampledata} along with the gradient-specific retention times.

\begin{table}
\setlength{\tabcolsep}{8 pt}
\centering
\caption{Sample of the ASO compounds and their observed $t_\mathrm{R}$ per each gradient time.}
\label{Sampledata}
\begin{tabular}{clccc} 
\hline
\vcell{\textbf{\#}}   & \vcell{\textbf{Sequence}} & \multicolumn{1}{l}{\vcell{\begin{tabular}[b]{@{}l@{}}\textbf{$t_\mathrm{R}$ (min)}\\\textbf{G1 (11 min)}\end{tabular}}} & \vcell{\textbf{G2 (22 min)}} & \vcell{\textbf{G3 (44 min)}}  \\[-\rowheight]
\printcelltop         & \printcelltop             & \multicolumn{1}{l}{\printcellmiddle}                                                                              & \printcellbottom             & \printcellbottom              \\ 
\hline
1                     & T*C*A*TTAGAA*T*T*A        & 7.901                                                                                                           & 11.113                     & 15.921                  \\
2                     & T*C*A*TTAGAA*T*T*A-P=O    & 7.869                                                                                                           & 11.031                  & 15.775                   \\
3                     & C*A*TTAGAA*T*T*A          & 7.624                                                                                                       & 10.510                  & 14.699                   \\
\multicolumn{1}{l}{4} & C*A*TTAGAA*T*T*A-P=O      & 7.608                                                                                                       & 10.444                  & 14.601                   \\
\multicolumn{1}{l}{5} & A*TTAGAA*T*T*A            & 7.298                                                                                                       & 9.841                  & 13.346                      \\
\multicolumn{1}{l}{6} & A*TTAGAA*T*T*A-P=O        & 7.282                                                                                                       & 9.711                  & 13.134                      \\
\hline
\end{tabular}
\end{table}

Each sequence of ASO compounds is a combination of four different nucleotide bases adenine (A), thymine (T), cytosine (C), and guanine (G). These bases form specific pairs (A with T and G with C), together forming a sequence that is an input for the ML model. The sequence could include other atoms, such as sulfur. In this dataset, sulfur atoms are denoted in *, forming partially or fully phosphorothioated sequences. The raw data consist of 876 unique phosphorothioated and non-phosphorothioated synthetic ASO sequences and their experimentally obtained retention times.

The dataset includes 8-, 12-, 16-, and 20-long ASO sequences. The compounds were chosen so that A, T, C, and G are randomly combined at different positions in the sequence. Of the 876 compounds, 79.6\% are partially or fully phosphorothioated with sulfur. A sequence ending with \emph{-P=O} denotes the loss of a sulfur atom at any position in the sequence. Regarding the nucleotide base in the first position, it is almost equally distributed among the four nucleotide bases. However, for the last position, 75.1\% have either A or C at the end of the sequence, with A dominating at 44.1\%.

\begin{table}
\setlength{\tabcolsep}{4 pt}
\centering
\caption{Percentages A, T, C, and G in the first and last positions in ASO sequences.}
\label{positiondata}
\begin{tabular}{ccc} 
\hline
\textbf{Nucleotide base} & \begin{tabular}[c]{@{}c@{}}\textbf{\% of nucleotide base}\\\textbf{in first position}\end{tabular} & \begin{tabular}[c]{@{}c@{}}\textbf{\% of nucleotide base}\\\textbf{in Last position}\end{tabular}  \\
\hline
A               & 24.4\%                                                                                    & 44.1\%                                                                                     \\
T               & 27.9\%                                                                                     & 14.7\%                                                                                     \\
C               & 23.3\%                                                                                     & 31.0\%                                                                                     \\
G               & 24.1\%                                                                                     & 10.0\%                                                                                     \\
\hline
\end{tabular}
\end{table}

\begin{figure}
  \centering
  \begin{subfigure}[b]{0.55\textwidth}
         \centering
         \includegraphics[width=\textwidth]{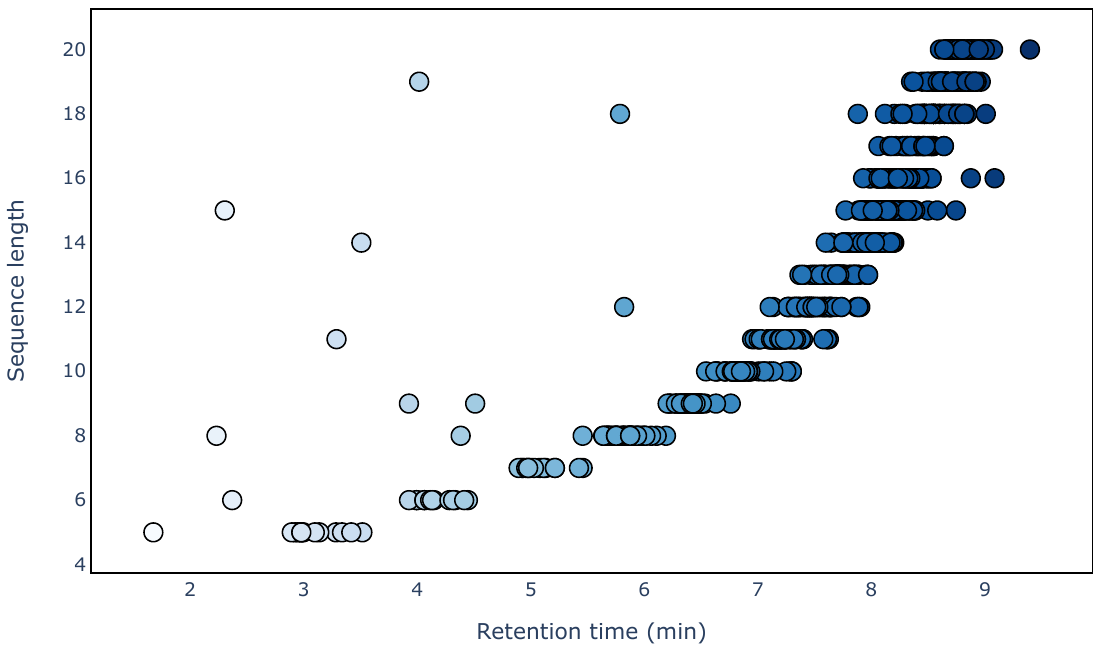}
         \caption{Gradient 1 (11 min)}
     \end{subfigure}
     \begin{subfigure}[b]{0.55\textwidth}
         \centering
         \includegraphics[width=\textwidth]{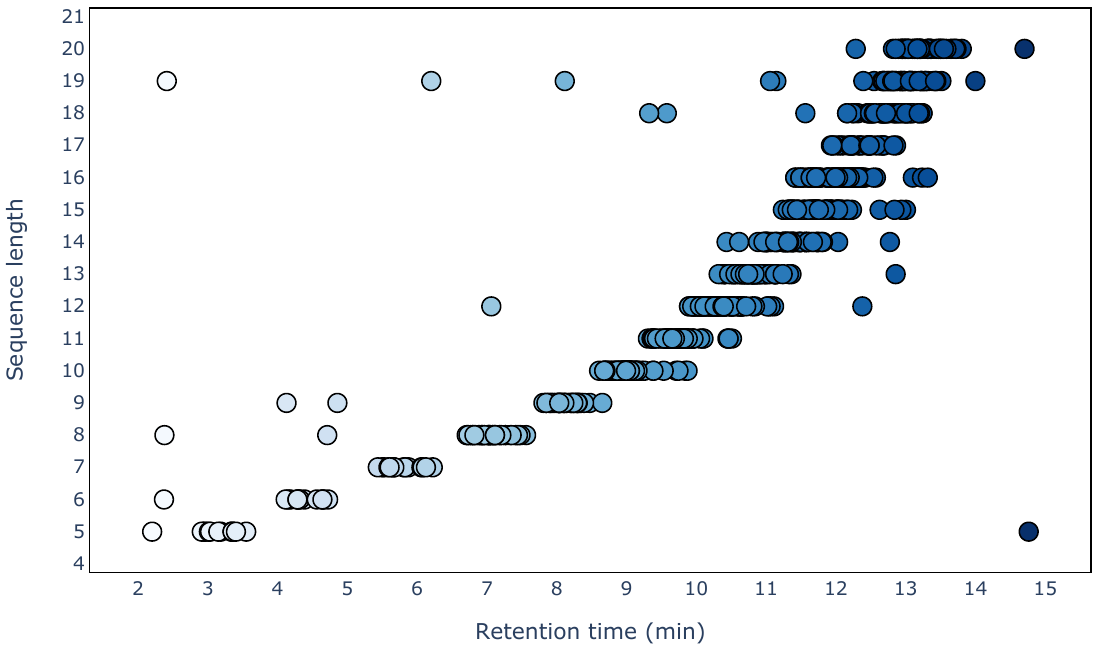}
         \caption{Gradient 2 (22 min)}
     \end{subfigure}
      \begin{subfigure}[b]{0.55\textwidth}
         \centering
         \includegraphics[width=\textwidth]{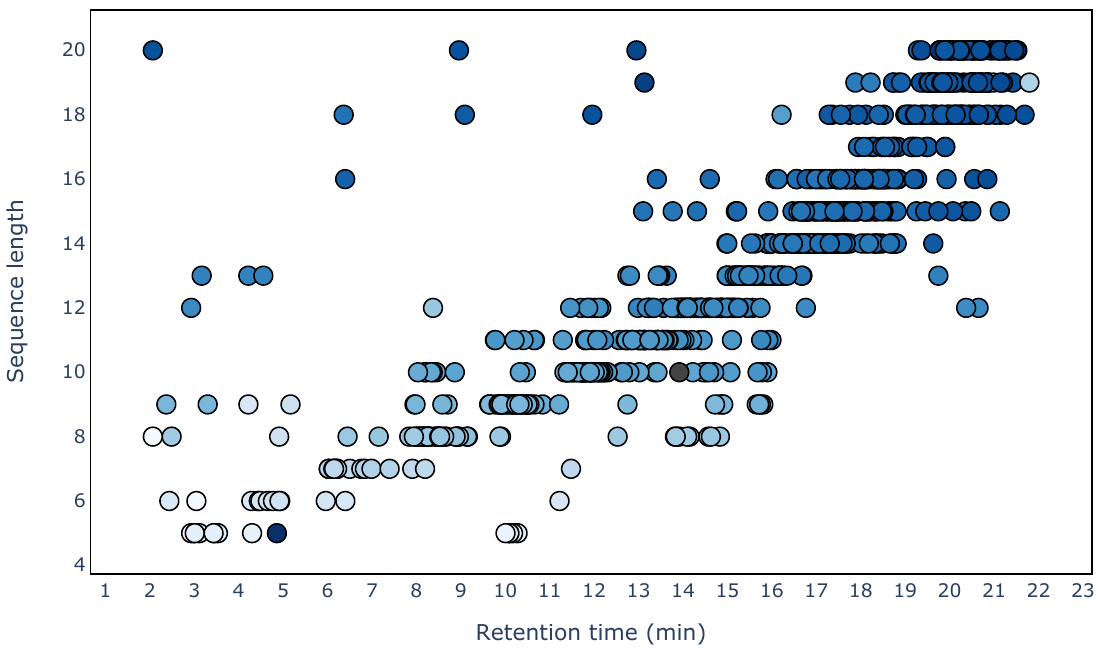}
         \caption{Gradient 3 (44 min)}
     \end{subfigure}
  \caption{Distribution of $t_\mathrm{R}$ across G1, G2, and G3}\label{fig:tRG123}
\end{figure}

The range of observed $t_\mathrm{R}$ differs among the three gradient times. The higher the gradient, the slower a compound elutes. Therefore, the longer a compound can be retained in an IPC system. The maximum recorded $t_\mathrm{R}$ for G1, G2, and G3 is 9.395, 14.762, and 21.795 min, respectively. In fact, most compounds elute in a specific time window, as shown in Figure \ref{fig:tRG123}.

\subsection{Feature Extraction} 

The feature selection process is crucial to the performance of ML models. Fortunately, the composition of the ASOs sequences allows various features to be derived. In this paper, the iterative process combined with domain knowledge expertise allowed stepwise experimentation and a selection of features to be included. Table \ref{features_tb} below describes the characteristics that were derived from the input data and included in the comparison analysis as independent variables. For simpler reporting and analysis, the derived features were grouped into five categories: COUNT, CONTACT, SCONTACT, SuCOUNT, and POSITION. The COUNT group includes four features that represent the frequency of each nucleotide and the total length of the sequence. CONTACT group includes the ordered occurrences of di-nucleotides, a total of 16 features. SCONTACT is similar to the CONTACT group but captures the frequencies of the unordered di-nucleotides and has six features. SuCOUNT represents the total number of sulfur atoms in a sequence. Finally, the POSITION group identifies the nucleotides in the first and last positions of a sequence. The first three groups of features were inherited from a previous study \cite{Strum2007}. In this paper, the POSITION and the SuCOUNT features were introduced as new groups to be investigated, and their impact on the models' predictive power was analyzed. In total, 30 features were derived from the compounds' sequences, forming the ML input data from all three datasets.

In ML problems, the features strongly influencing the target variable are often unknown. Therefore, testing all possible combinations of features is necessary to realize the best estimator. In previous research, the COUNT group of features was recognized as highly influential in predicting $t_\mathrm{R}$ \cite{ENMARK2022}. However, other feature groups were shown to be less influential and only increase computational time. This paper tests all 31 possible combinations of feature groups and outputs the best-optimized estimator combined with the optimal feature selection.

\begin{table}
\centering
\caption{Overview of the derived groups of characteristics and description.}
\label{features_tb}
\setlength{\tabcolsep}{8 pt}
\begin{tabular}{lcl} 
\hline
\textbf{Feature group: number}     & \textbf{Number of features} & \textbf{Feature description}                                                                                                                                                                                \\ 
\hline
\textbf{COUNT: 1-4}                & 4 features                  & Frequencies of nucleotide bases A, T, C, and G in a sequence                                                                                                                                               \\
\textbf{COUNT: 5}                  & 1 feature                   & Length of a sequence                                                                                                                                                                                        \\
\vcell{\textbf{CONTACT: 6 - 21}}   & \vcell{16 features}         & \vcell{\begin{tabular}[b]{@{}l@{}}Frequency of occurrences of ordered di-nucleotides i.e., \\frequency of occurrence of~AA, AT, AC, AG, TA, TT, TG, \\TC, CA, CT, CG, CC, GA, GT, GG, and GC.\end{tabular}}  \\[-\rowheight]
\printcelltop                      & \printcelltop               & \printcellmiddle                                                                                                                                                                                            \\
\vcell{\textbf{SCONTACT: 22 - 27}} & \vcell{6 features}          & \vcell{\begin{tabular}[b]{@{}l@{}}Frequency of occurrences of unordered di-nucleotides \\i.e the occurrence of~AT\_TA, AG\_GA, AC\_CA, \\{}GT\_TG, CT\_TC, CG\_GC\end{tabular}}                 \\[-\rowheight]
\printcelltop                      & \printcelltop               & \printcellmiddle                                                                                                                                                                                            \\
\textbf{SuCOUNT: 28}               & 1 feature                   & Number of sulfur atoms in a sequence                                                                                                                                                                        \\
\textbf{POSITION: 29}              & 1 feature                   & Nucleotide base in the first position                                                                                                                                                                       \\
\textbf{POSITION: 30}              & 1 feature                   & Nucleotide base in the last position                                                                                                                                                                        \\
\hline
\end{tabular}
\end{table}

\begin{table}
\centering
\caption{A sample of the encoded data fed to ML models as input data.}
\label{tbl:encoded_features}
\resizebox{\textwidth}{!}{
\begin{tabular}{cccccccccccccccccccc}
\toprule
\textbf{Sequence} & \textbf{Fa} & \textbf{Ft} & \textbf{Fg} & \textbf{Fc} & \textbf{Length} & \textbf{AT} & \textbf{AC} & \textbf{TA} & \textbf{TG} & \textbf{GA} & \textbf{CT} & \textbf{AT\_TA} & \textbf{AC\_CA} & \textbf{TG\_GT} & \textbf{GA\_AG} & \textbf{CT\_TC} & \textbf{Fsulf} & \textbf{First} & \textbf{Last} \\
\midrule
TG*ACTATG & 2  & 3  & 2  & 1  & 8      & 1  & 1  & 1  & 2  & 1  & 1  & 2     & 1     & 2     & 1     & 1     & 2     & 1     & 3    \\
\bottomrule
\end{tabular}
}
\end{table}

\subsection{Motivation for the selection of models}
As part of this paper, several ML models were selected and evaluated in the three datasets. The careful selection of models resulted in the choice of Random Forest (RF), Gradient Boosting (GB), Decision Tree (DT), and SVR algorithms.

Since being introduced in 2001 \cite{Breiman2001}, the RF algorithm has proven to be a successful algorithm for classification and regression problems \cite{ESTEVE2023}. The algorithm is based on randomly selecting input samples from the training set to build the predictor trees. In addition to bagging, another aspect of randomization is introduced through the randomized selection of features at each node \cite{Dai2022}. Then, the aggregated prediction decision is made by averaging the predictions of the individual trees \cite{ESTEVE2023}. RF is an ensemble-based algorithm built to reduce overfitting and increase accuracy \cite{CHUNG2023}. Given its work method, RF is considered a relatively robust, simple, and fast algorithm for outliers. The RF algorithm is described in detail in \cite{Breiman2001}.

The second selected algorithm is GB. GB is another ensemble-based algorithm with high prediction accuracy compared to other models \cite{CHUNG2023}. Boosting is an approach that improves the predictive power of a learning algorithm by iteratively combining weak learners \cite{Bentejac2021}. This iterative process of increasing the weights of the wrongly classified examples results in a relatively accurate model \cite{Bentejac2021}. GB is one form of regression tree boosting, which is based on gradient descent minimization \cite{CHUNG2023}. 

Regarding the DT algorithm, it follows the divide-and-conquer strategy to represent the learned knowledge, mimicking human reasoning \cite{Bentejac2021}. The advantage of the DT algorithm comes from its ability to break down complex decisions into explainable ones such that the selection rules are represented in the branches of the tree \cite{Bukhari2022}. The algorithm is considered fast and efficient with small datasets and results in a group of rules used in the prediction of unseen data \cite{MAHOTO2023}. In fact, interpretability is considered one of the most significant advantages of single-tree models \cite{Costa2023}. Since smaller trees are less prone to overfitting, pre-prune and post-prune methods are used to enhance generalizability \cite{YUAN2021}. In this sense, the goal is to construct the simplest possible tree with the lowest error value. 

Finally, SVM was selected because it offers an alternative to other learning methods that aim to minimize the mean squared error. The novelty of SVM comes from the so-called kernel trick, which enables efficient execution of the algorithm \cite{VapnikSpam1999}. The SVM method finds two hyper-planes in the high-dimensional space that best separate the data into two classes while maximizing the margin between them \cite{Vapnik1999}. An advantage of SVM is that it is relatively insensitive to the size of the training examples in each class \cite{VapnikSpam1999}. This characteristic motivated researchers to use this method dominantly in $t_\mathrm{R}$ estimation. However, it should be noted that the training time of the SVM algorithms can increase considerably in proportion to the size of the training examples \cite{VapnikSpam1999}. The SVM is famously used in classification problems but can also be applied to regression problems \cite{Kurani2023}, as in this case study. Then, it is called SVR.

\subsection{Training and evaluating ML models}
\begin{figure}
\centering
         \centering
         \includegraphics[scale=.7]{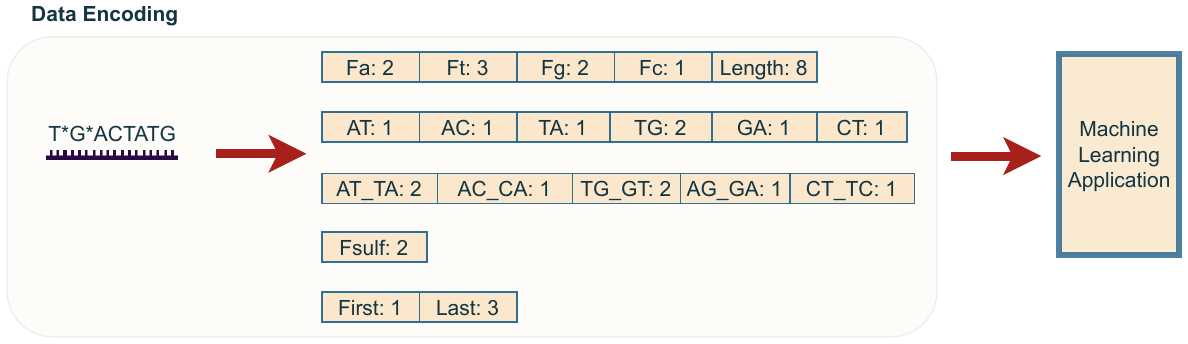}
         \caption{Encoding the composition characteristics of an ASO sequence into quantitative features in preparation for the application of ML models. }\label{fig:featencoding}
\end{figure}

Raw data are rarely numeric; they are usually collected as a combination of different types, such as nominal, ordinal, numerical, images, and text. However, several popular ML models can only work with numeric quantities \cite{MATOS2022}. The model can then deduce the relations among these numerical quantities and learn something useful about the data. In this paper, 30 characteristics were derived from the composition of the compounds and transformed into a comprehensible format by the ML models. Figure \ref{fig:featencoding} shows an example of encoding 'T*G*ACTATG' into values that the ML model understands. Fa, Ft, Fg, and Fc represent the frequency of occurrence of each of the nucleotide bases in the sequence. For example, in the sequence 'T*G*ACTATG', the adenine nucleotide base occurs 2 times. The second group of features represents the ordered occurrence of dinucleotides, where in this example, the AT dinucleotide occurs only once in the sequence. Whereas, the unordered occurrence of AT\_TA is observed twice in the sequence. Two sulfur atoms are present and represented in the Fsulf frequency. The 'First' and the 'Last' features, representing the first and last positions in the sequences, are encoded such that A:0, T:1, C:2, G:3. In Figure \ref{fig:featencoding}, only the features with non-zero values are shown. As a result of the encoding pre-processing task, the data fed to the ML model includes a dataframe of the encoded numerical representation of each sequence in the form of columns. It should be noted that the feature derivation and encoding process has been applied similarly to the train and test datasets.

For every gradient dataset, the data is divided into 80\% train set and the remaining 20\% is a test set. The test set is never used for developing the ML model, so that part of the data is kept independent from the learned distribution in the train set \cite{Karasiak2022}. This allows for fair evaluation and performance comparison among the selected models. For the purpose of a fair comparison, all models were evaluated on the same test set. 

Two essential optimization steps were performed for every selected model, involving hyperparameter tuning and optimal feature selection. The two steps were combined, where a three-fold cross-validation grid search was performed for each unique combination of features. The grid search results were evaluated, and the optimal parameters and feature groups per model were recorded. Table \ref{tbl:hypertuning} summarizes the different combinations of parameters in the grid that were tested to optimize the performance of the ML models, where the chosen parameters for each of the models are marked in bold. All ML applications, including the hyperparameter tuning, were performed using Python version 3.9.7 and supported by the scikit-learn version 0.24.2 and other relevant libraries.

\begin{table}
\caption{Hypertuning parameter grids for the different ML models}
\label{tbl:hypertuning}
\begin{tabular}{ll}
\hline
\textbf{Model} & \textbf{Parameter Grid} \\
\hline
GB & \parbox{6cm}{
\begin{itemize}
\item 'max\_depth': [5, 10, 15, \textbf{20}, 50]
\item 'learning\_rate': [0.001, \textbf{0.01}, 0.1, 0.2]
\item 'n\_estimators': [100, 500, \textbf{1000}]
\item 'max\_leaf\_nodes': [\textbf{2}, 5, 10]
\end{itemize}
} \\
\hline
RF & \parbox{6cm}{
\begin{itemize}
\item 'n\_estimators': [int(x) for x in [ \textbf{10}, 82.5, 155, 227.5, 300]]
\item 'max\_depth': [int(x) for x in [\textbf{5}, 60, 115, ... 500]]
\item 'min\_samples\_split': [2, \textbf{4}]
\item 'min\_samples\_leaf': [\textbf{1}, 2]
\end{itemize}
} \\
\hline
DT & \parbox{6cm}{
\begin{itemize}
\item 'max\_depth': [2, 10, \textbf{20}, 30, 40, 100]
\item 'min\_samples\_split': [2, \textbf{4}, 8]
\item 'min\_samples\_leaf': [1, 2, 4, 6, \textbf{8}]
\end{itemize}
} \\
\hline
SVR & \parbox{6cm}{
\begin{itemize}
\item 'C': [1., 21.3877551, \textbf{62.16326}, ...1000.]
\item 'gamma': [0.1, \textbf{0.01}, 0.001]
\item 'epsilon':[0.0001, \textbf{0.00011111}, 0.00012222, 0.0002]
\item 'kernel': \textbf{'rbf'}
\end{itemize}
} \\
\hline
\end{tabular}
\end{table}

The grid search method used aims to minimize RMSE as an objective function, which refers to the average error between the observed and predicted $t_\mathrm{R}$ values. RMSE is an absolute error measurement that is commonly used for model fitting, selection, and comparisons with other models \cite{Karunasingha2022}. RMSE is given in the following equation:
\begin{equation}
RMSE = \sqrt{\frac{1}{n}\sum_{i=1}^{n}(y_i - \hat{y}_i)^2}
\end{equation}

\begin{figure}
\centering
         \centering
         \includegraphics[scale=.45]{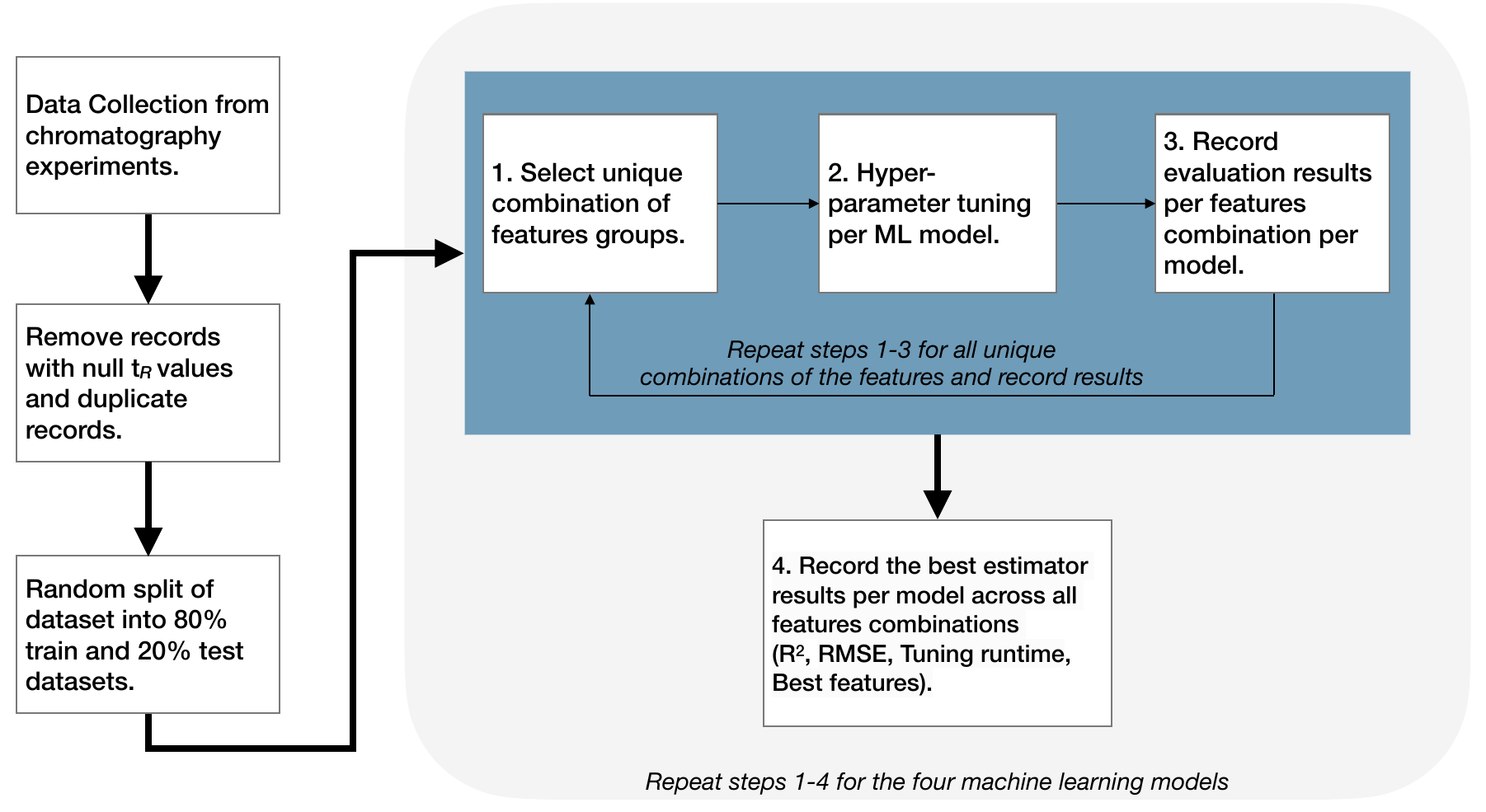}
         \caption{From data collection to ML step-by-step block diagram. Data preparation and ML phases are repeated for each of the three datasets.}
         \label{fig:blockdiag}
\end{figure}

RMSE is not the only metric observed in our paper. Other evaluation metrics, such as $R^2$ and run time, are also recorded. $R^2$ provides a useful indication of the explained variance in the fitted model \cite{Rights2023}. Figure \ref{fig:blockdiag} describes the steps performed from data preparation to the application of ML, which resulted in the selection of the best estimators.

The choice of hyperparameters for each model, except for the SVR, is based on observations made from previous trials with a smaller but very similar dataset. The choice of hyperparameters for the SVR model is inspired by a previous study by Enmark \textit{et al.} in which they tested the SVR model \cite{ENMARK2022}. However, the SVR hyperparameter space has been slightly adjusted for the purpose of achieving a feasible tuning run time.

\section{Results and Discussion}\label{sec:results}
\begin{figure*}
\centering
    \begin{subfigure}[b]{0.23\textwidth}
         \centering
         \includegraphics[width=\textwidth]{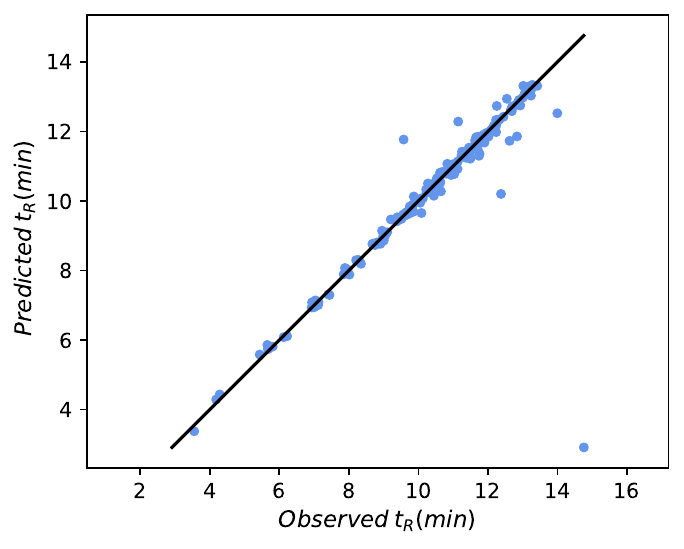}
         \caption{SVR}
     \end{subfigure}
    \begin{subfigure}[b]{0.23\textwidth}
         \centering
         \includegraphics[width=\textwidth]{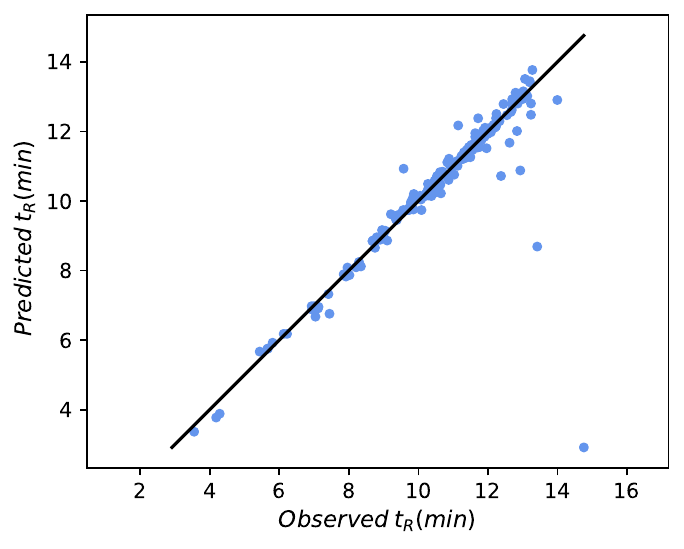}
         \caption{Gradient boost}
     \end{subfigure}
     \begin{subfigure}[b]{0.23\textwidth}
         \centering
         \includegraphics[width=\textwidth]{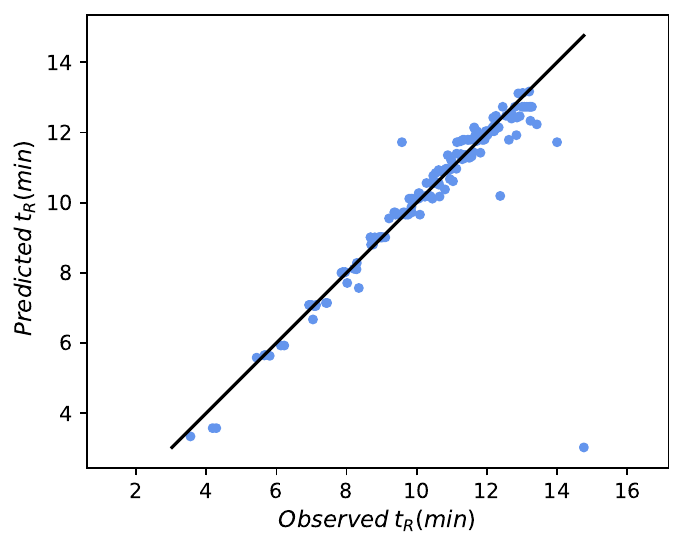}
         \caption{Random Forest}
     \end{subfigure}
     \begin{subfigure}[b]{0.23\textwidth}
         \centering
         \includegraphics[width=\textwidth]{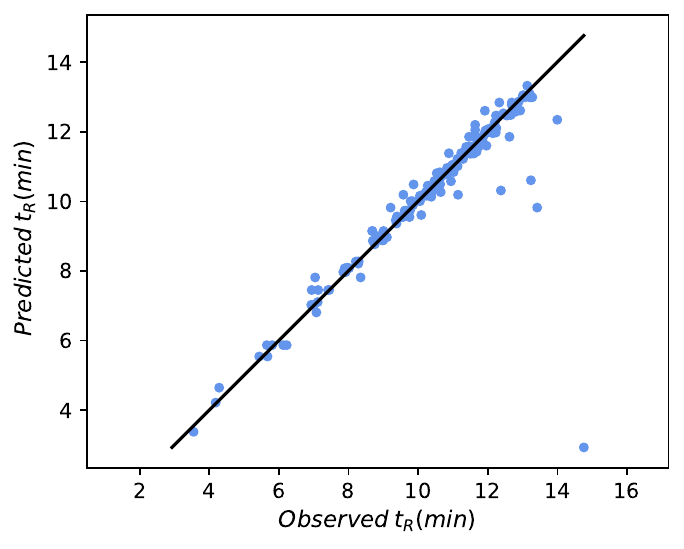}
         \caption{Decision tree}
     \end{subfigure}
    \caption{Observed versus predicted $t_\mathrm{R}$ for G2 test dataset.}
    \label{img:G2res}
\end{figure*}

\begin{figure}
\centering
\begin{subfigure}[b]{0.23\textwidth}
         \centering
         \includegraphics[width=\textwidth]{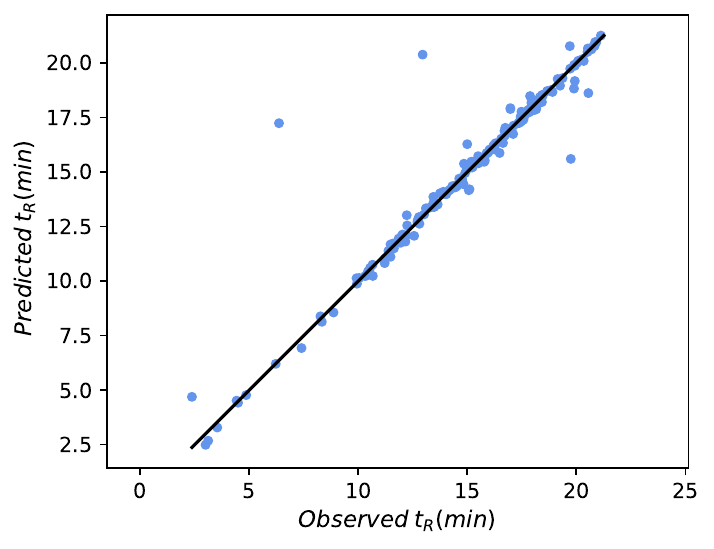}
         \caption{SVR}
     \end{subfigure}
    \begin{subfigure}[b]{0.23\textwidth}
         \centering
         \includegraphics[width=\textwidth]{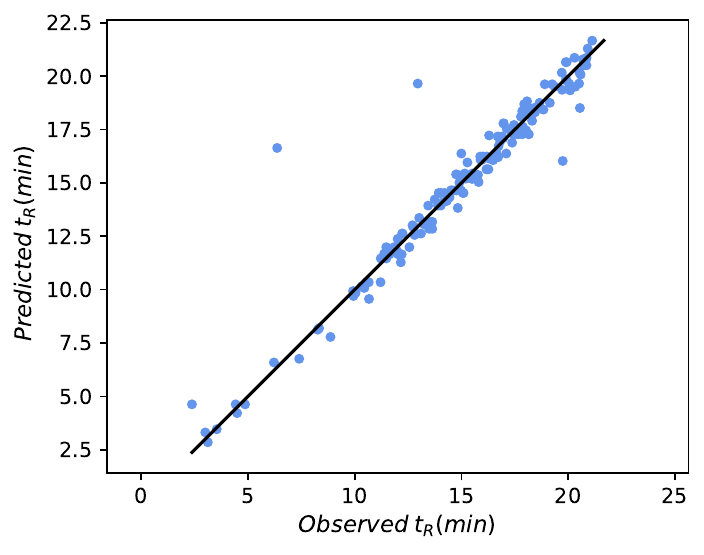}
         \caption{Gradient boost}
     \end{subfigure}
     \begin{subfigure}[b]{0.23\textwidth}
         \centering
         \includegraphics[width=\textwidth]{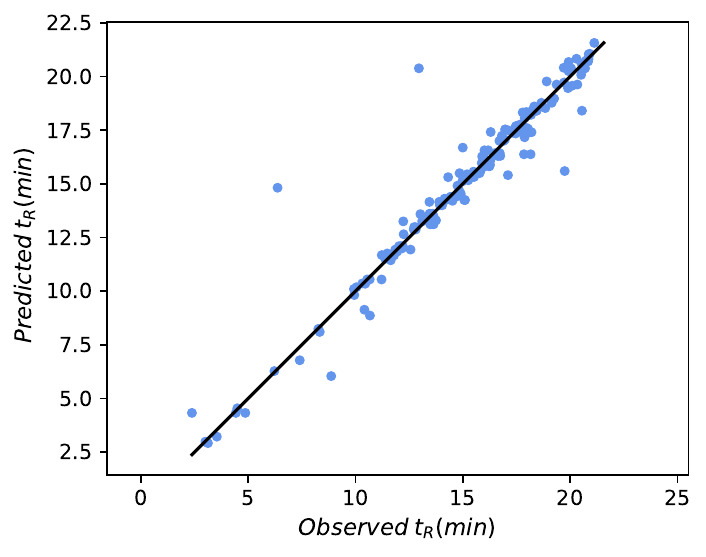}
         \caption{Random Forest}
     \end{subfigure}
     \begin{subfigure}[b]{0.23\textwidth}
         \centering
         \includegraphics[width=\textwidth]{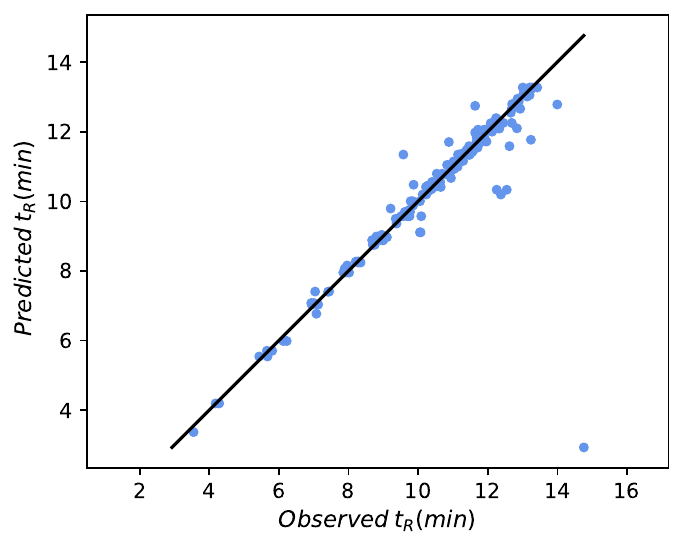}
         \caption{Decision tree}
     \end{subfigure}
    \caption{Observed versus predicted $t_\mathrm{R}$ for G3 test dataset.}
    \label{img:G3res}
\end{figure}

Four ML models, SVR, GB, RF, and DT, were trained and evaluated across three ASO datasets (G1, G2, G3). To evaluate the performance of these models, several criteria were selected and reported in Table \ref{tab:tbresults}. Figures \ref{img:G1res}, \ref{img:G2res}, and \ref{img:G3res} visualize the observed versus predicted $t_\mathrm{R}$ for the ML models tested in G1, G2 and G3, respectively. Accuracy criteria are reported as RMSE training and test, and $R^2$. In addition, the hyperparameter tuning run time is recorded in minutes. Finally, the most influential predictors selected by the best estimator in each model are noted. The designed metrics grid allows us to compare the performance of ML models in terms of accuracy, speed, and complexity.

\begin{table}
\centering
\setlength{\tabcolsep}{2 pt}
\caption{Evaluation of the performance of ML models in train and test datasets for the three gradients.}\label{tab:tbresults}
\resizebox{\textwidth}{!}{
\begin{tabular}{cccccccl} 
\hline
\vcell{\textbf{Gradient}}   & \vcell{\textbf{ML model}} & \vcell{\textbf{RMSE train}} & \vcell{\textbf{R\textsuperscript{2} train}} & \vcell{\textbf{RMSE test}} & \vcell{\textbf{R\textsuperscript{2} test}} & \vcell{\textbf{Run time (min)}} & \vcell{\textbf{Best estimator's features}}  \\[-\rowheight]
\printcelltop               & \printcelltop             & \printcelltop               & \printcelltop                               & \printcelltop              & \printcelltop                              & \printcelltop                   & \printcelltop                               \\ 
\hline
\multirow{4}{*}{Gradient 1} & \vcell{\textbf{ SVR }}    & \vcell{0.334}              & \vcell{0.912}                               & \vcell{0.442}              & \vcell{0.861}                              & \vcell{404.443}                 & \vcell{SCONTACT, POSITION}                  \\[-\rowheight]
                            & \printcelltop             & \printcelltop               & \printcelltop                               & \printcelltop              & \printcelltop                              & \printcelltop                   & \printcelltop                               \\
                            & \vcell{\textbf{ RF }}     & \vcell{0.295}               & \vcell{0.94}                                & \vcell{0.508}              & \vcell{0.772}                              & \vcell{52.570}                  & \vcell{CONTACT, SCONTACT}                   \\[-\rowheight]
                            & \printcelltop             & \printcelltop               & \printcelltop                               & \printcelltop              & \printcelltop                              & \printcelltop                   & \printcelltop                               \\
                            & \vcell{\textbf{ GB }}     & \vcell{0.368}               & \vcell{0.906}                               & \vcell{0.439}              & \vcell{0.830}                              & \vcell{47.606}                  & \vcell{COUNT, SCONTACT, POSITION}           \\[-\rowheight]
                            & \printcelltop             & \printcelltop               & \printcelltop                               & \printcelltop              & \printcelltop                              & \printcelltop                   & \printcelltop                               \\
                            & \vcell{\textbf{ DT }}     & \vcell{0.29}                & \vcell{0.94}                                & \vcell{0.501}              & \vcell{0.778}                              & \vcell{0.345}                   & \vcell{COUNT, CONTACT, POSITION}            \\[-\rowheight]
                            & \printcelltop             & \printcelltop               & \printcelltop                               & \printcelltop              & \printcelltop                              & \printcelltop                   & \printcelltop                               \\ 
\hline
\multirow{4}{*}{Gradient 2} & \vcell{\textbf{ SVR }}    & \vcell{0.641}               & \vcell{0.917}                               & \vcell{0.953}            & \vcell{0.783}                              & \vcell{223.26}                  & \vcell{COUNT, SuCOUNT}                      \\[-\rowheight]
                            & \printcelltop             & \printcelltop               & \printcelltop                               & \printcelltop              & \printcelltop                              & \printcelltop                   & \printcelltop                               \\
                            & \vcell{\textbf{ RF }}     & \vcell{0.586}               & \vcell{0.93}                                & \vcell{0.977}              & \vcell{0.772}                              & \vcell{53.206}                  & \vcell{COUNT, SuCOUNT, SCONTACT}            \\[-\rowheight]
                            & \printcelltop             & \printcelltop               & \printcelltop                               & \printcelltop              & \printcelltop                              & \printcelltop                   & \printcelltop                               \\
                            & \vcell{\textbf{ GB }}     & \vcell{0.316}               & \vcell{0.98}                                & \vcell{1.021}              & \vcell{0.75}                               & \vcell{47.565}                  & \vcell{COUNT, SuCOUNT, CONTACT, POSITION}   \\[-\rowheight]
                            & \printcelltop             & \printcelltop               & \printcelltop                               & \printcelltop              & \printcelltop                              & \printcelltop                   & \printcelltop                               \\
                            & \vcell{\textbf{ DT }}     & \vcell{0.505}               & \vcell{0.949}                               & \vcell{0.986}              & \vcell{0.767}                              & \vcell{0.345}                   & \vcell{COUNT, POSITION}                     \\[-\rowheight]
                            & \printcelltop             & \printcelltop               & \printcelltop                               & \printcelltop              & \printcelltop                              & \printcelltop                   & \printcelltop                               \\ 
\hline
\multirow{4}{*}{Gradient 3} & \vcell{\textbf{ SVR }}    & \vcell{1.0}                 & \vcell{0.92}                                & \vcell{1.08}               & \vcell{0.925}                              & \vcell{177.494}                 & \vcell{COUNT, SuCOUNT, CONTACT}             \\[-\rowheight]
                            & \printcelltop             & \printcelltop               & \printcelltop                               & \printcelltop              & \printcelltop                              & \printcelltop                   & \printcelltop                               \\
                            & \vcell{\textbf{ RF }}     & \vcell{1.035}               & \vcell{0.942}                               & \vcell{1.17}               & \vcell{0.928}                              & \vcell{73.284}                  & \vcell{COUNT, CONTACT, SCONTACT}            \\[-\rowheight]
                            & \printcelltop             & \printcelltop               & \printcelltop                               & \printcelltop              & \printcelltop                              & \printcelltop                   & \printcelltop                               \\
                            & \vcell{\textbf{ GB }}     & \vcell{1.1}                 & \vcell{0.926}                               & \vcell{1.18}               & \vcell{0.925}                              & \vcell{108.84}                  & \vcell{COUNT, POSITION}                     \\[-\rowheight]
                            & \printcelltop             & \printcelltop               & \printcelltop                               & \printcelltop              & \printcelltop                              & \printcelltop                   & \printcelltop                               \\
                            & \vcell{\textbf{ DT }}     & \vcell{0.505}               & \vcell{0.949}                               & \vcell{0.986}              & \vcell{0.767}                              & \vcell{0.345}                   & \vcell{COUNT, POSITION}                     \\[-\rowheight]
                            & \printcelltop             & \printcelltop               & \printcelltop                               & \printcelltop              & \printcelltop                              & \printcelltop                   & \printcelltop                               \\
\hline
\end{tabular}
}
\end{table}

Overall, the best estimators across all gradients relied mainly on the length of the sequences and the frequencies of nucleotide bases, in addition to their position. Hence, it is concluded that the features in the COUNT group positively influenced the predictability task. SVR is the least sensitive model to the variations in data points, and RF is the most sensitive. The flexibility of a model is important to achieve accurate predictions, but increased flexibility also leads to learning noise. Furthermore, newly suggested features representing the sulfur count and the nucleotides residing at the first and last positions of a sequence have shown to improve the predictive power of the models, as seen in the G1, G2, and G3 results. 

The GB model was tuned in 68.0 min compared to the SVR, which, on average, required 4.47 hours. The average tuning run time for the RF model was 59.68 and 0.34 min for the DT model. The DT model was the fastest among all three models, with a remarkable difference in run time, but yielded the highest error across all three datasets due to overfitting.

\subsection{Performance on the G1 dataset}

\begin{figure}
\centering
     \begin{subfigure}[b]{0.23\textwidth}
         \centering
         \includegraphics[width=\textwidth]{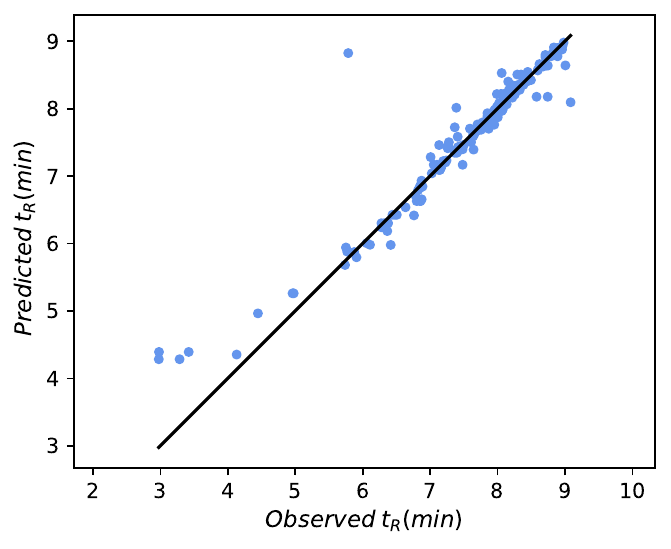}
         \caption{SVR}
     \end{subfigure}
    \begin{subfigure}[b]{0.23\textwidth}
         \centering
         \includegraphics[width=\textwidth]{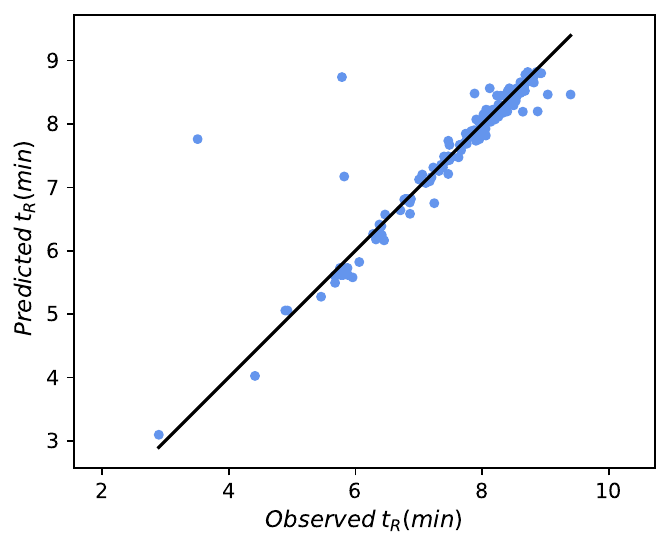}
         \caption{Gradient boost}
     \end{subfigure}
     \begin{subfigure}[b]{0.23\textwidth}
         \centering
         \includegraphics[width=\textwidth]{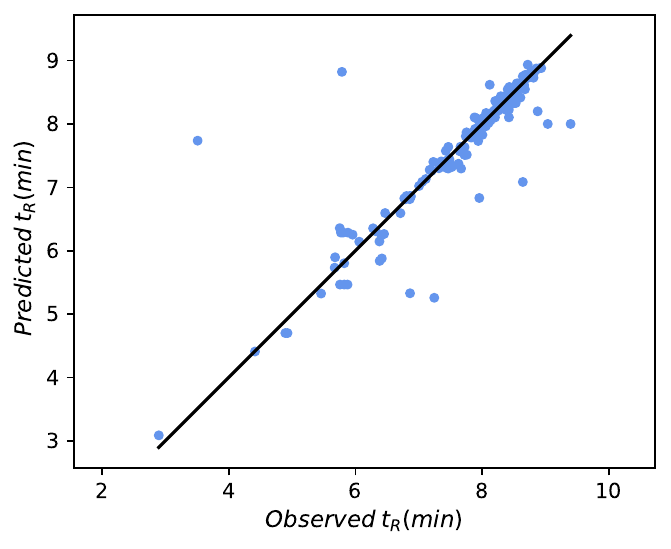}
         \caption{Random Forest}
     \end{subfigure}
     \begin{subfigure}[b]{0.23\textwidth}
         \centering
         \includegraphics[width=\textwidth]{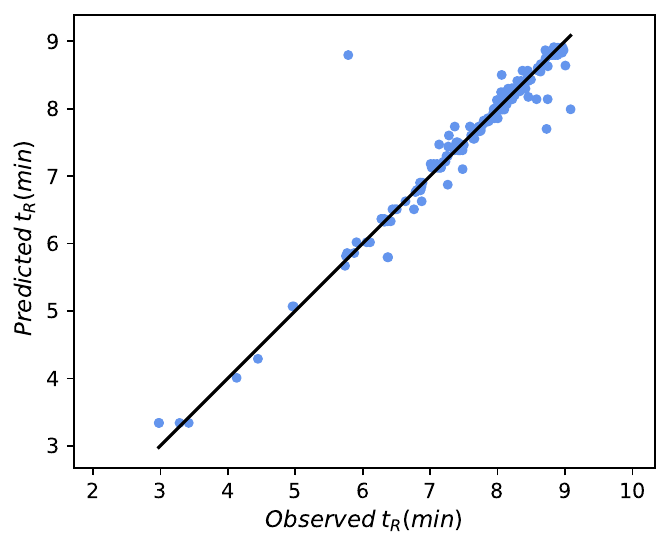}
         \caption{Decision tree}
     \end{subfigure}
    \caption{Observed versus predicted $t_\mathrm{R}$ for G1 test dataset.}
    \label{img:G1res}
\end{figure}

In terms of RMSE, the models performed best in the G1 dataset. This is expected given that the G1 data are the least noisy among the three datasets.  In the G1 dataset, the performance of SVR and GB was relatively close, with GB performing slightly better than SVR as shown in Figure \ref{img:G1res}. GB achieved an RMSE test of 0.439 min in G1, while SVR recorded 0.442 min. SVR utilized 8 of the 30 features to achieve the best result, while GB needed an additional 5 features to achieve a similar result. However, the hyperparameter tuning of the SVR model was 8.49 times slower than GB. On average, tuning the SVR was the most computationally expensive, recording the slowest run time in the G2 dataset at 6.74 hours. SCONTACT and POSITION, the newly introduced features, were identified together or individually as important features for achieving the lowest average error. This shows that the arrangement of di-nucleotides has a certain influence on the prediction of $t_\mathrm{R}$. In addition to SCONTACT and POSITION, GB utilized the COUNT group of features to achieve the lowest RMSE. COUNT group is expected to be of great importance, as proven in \citep{ENMARK2022}.

\subsection{Performance on the G2 dataset}

For the G2 dataset, SVR and RF competed for the first and second best performing models. SVR outperformed RF with RMSE test of 0.953 min. However, the RF hyperparameter tuning was 4.16 times faster and produced results in 53.2 min. In this dataset, COUNT and SuCOUNT were shown to be influential in achieving the best results. In addition to both groups, RF also identified the SCONTACT group as important.

\subsection{Performance on the G3 dataset}

The four models were also trained and tested in the G3 dataset, and RF showed the best performance. All models performed poorly in this dataset compared to their performance in the G1 dataset. This is due to the volatility and the experimental noise reflected in the G3 data. The noisy nature of this dataset also influenced the hyperparameter optimization time, as it significantly increased the average time for RF and GB by 1.81 times. GB was always the fastest across all gradients. However, SVR recorded the lowest tuning run time compared to G2 and G1 despite the sensitive nature of the data in G3. But, SVR remained computationally expensive compared to the other three models. As the best-performing model on this dataset, SVR identified that the COUNT, SuCOUNT, and CONTACT features are all important. This suggests that the ordered arrangement of di-nucleotides can impact the prediction error.

\subsection{G3 In-Depth Error Analysis}
\begin{figure}
\centering
    \begin{subfigure}{0.4\textwidth}
         \includegraphics[width=\textwidth]{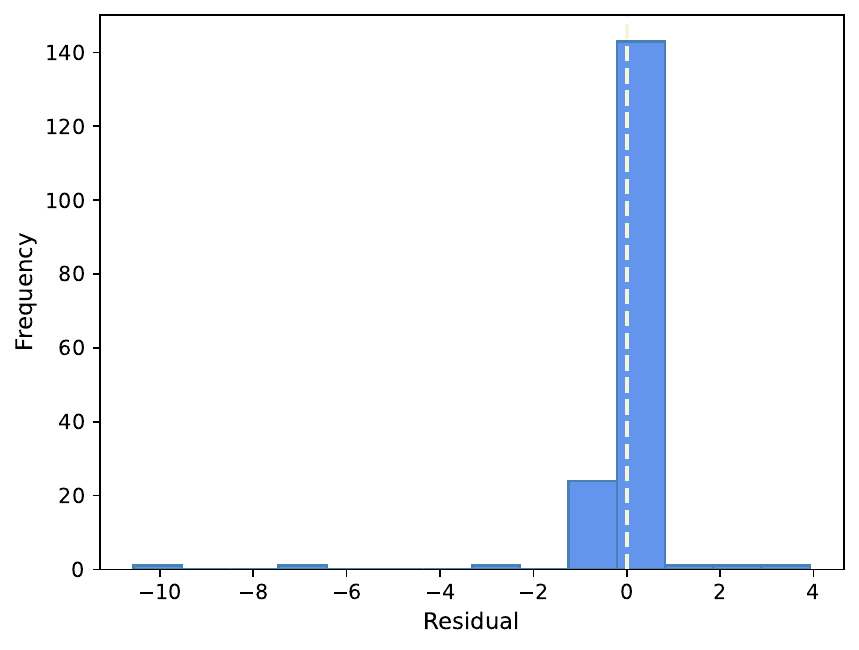}
         \caption{Histogram showing the frequency of overestimated and underestimated predictions.}
         \label{fig:SVREM}
     \end{subfigure}
     \begin{subfigure}{0.4\textwidth}
         \includegraphics[width=\textwidth]{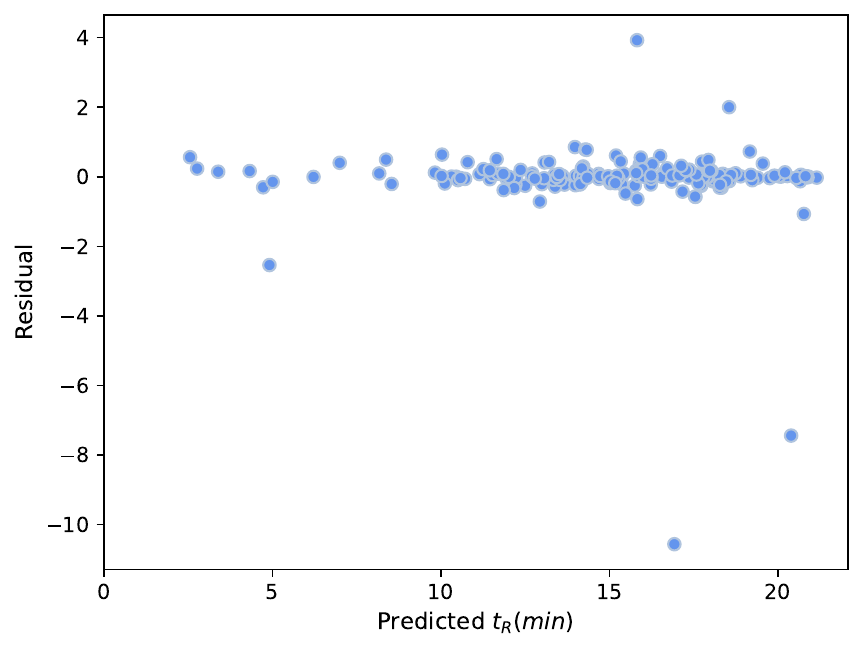}
         \caption{Residual plot showing the difference (min) between the observed and the predicted $t_\mathrm{R}$}
         \label{fig:SVRRP}
     \end{subfigure}
        \caption{Error Analysis plots referring to the test results of the optimized SVR model on G3 dataset.}
        \label{fig:ResidualPlot}
\end{figure}

The poor performance of the different ML models in G3, compared to G1 and G2 datasets, is explained by the noisiness and unpredictable nature of this dataset. Therefore, an in-depth error analysis is performed on the G3 dataset to understand the least predictable compounds. The results of the analysis are shown in Table \ref{tbl:residuals}, \ref{tbl:RFresiduals}, and \ref{tbl:GBresiduals}. The G3 test dataset includes 173 unique ASO compounds. In total, 95 predicted values of $t_\mathrm{R}$ were underestimated by the best-performing SVR model, and 78 predictions were overestimated. The histogram in Figure \ref{fig:ResidualPlot} shows the tendency of the SVR model to underestimate $t_\mathrm{R}$. The residual plot in Figure \ref{fig:ResidualPlot} shows the predicted values on the x-axis and the residual values on the y-axis. The residual values are calculated by subtracting the predicted $t_\mathrm{R}$ values from the observed $t_\mathrm{R}$ values. Positive residual values represent underestimated predictions, and negative residual values represent overly estimated predictions values.

Based on the error analysis visualizations in Figure \ref{fig:ResidualPlot}, the anomalies that had the highest residuals, whether they were over- or underestimated, are described in Table \ref{tbl:residuals}. The sequences listed in Table \ref{tbl:residuals} had residual values greater than one or less than -1. Table \ref{tbl:residuals} shows that all ASO sequences with large residuals have lost a sulfur atom, which is represented by \emph{(-P=O)} at the end of the sequence. The majority of the ASO sequences are 20 nucleotide long and fully phosphorothioated.

\begin{table}
\setlength{\tabcolsep}{2 pt}
\caption{ ASO sequences with the highest residual values in the G3 dataset. SVR predictions are labeled as over- or underestimated based on the calculated negative or positive values of residuals.}\label{tbl:residuals}
\begin{tabular}{clcccc} 
\hline
\textbf{\#} & \textbf{Sequence} & \multicolumn{1}{c}{\textbf{Observed $t_\mathrm{R}$}} & \multicolumn{1}{c}{\textbf{Predicted $t_\mathrm{R}$}} & \multicolumn{1}{c}{\textbf{Residual}} & \textbf{Over/underestimation}  \\ 
& & (min) & (min) & (min) & \\ 
\hline
1           & G*T*G*G*G*T*G*G*G*T*G*G*G*T*G*G*G*T-P=O   & 6.363                  & 16.927                  & -10.563        & overestimated                 \\
2           & A*T*C*A*G*T*A*T*T*A*A*A*A*T*T*T*T*C*A-P=O & 12.951                 & 20.392                  & -7.441         & overestimated                 \\
3           & A*C*T*A*T*G-P=O                           & 2.371                  & 4.909                   & -2.537         & overestimated                 \\
4           & T*T*T*T*T*T*T*T*T*T*T*T*T*T*T*T*T*T-P=O   & 19.701                 & 20.770                  & -1.069         & overestimated                 \\
5           & T*TTATCGGCGGG*A*A*C*A-P=O                 & 20.552                 & 18.551                  & 2.000          & underestimated                \\
6           & A*T*C*G*G*C*G*G*G*A*A*C*A-P=O             & 19.746                & 15.819                  & 3.926          & underestimated                \\
\hline
\end{tabular}
\end{table}

\begin{figure}
\centering
    \begin{subfigure}{0.4\textwidth}
         \includegraphics[width=\textwidth]{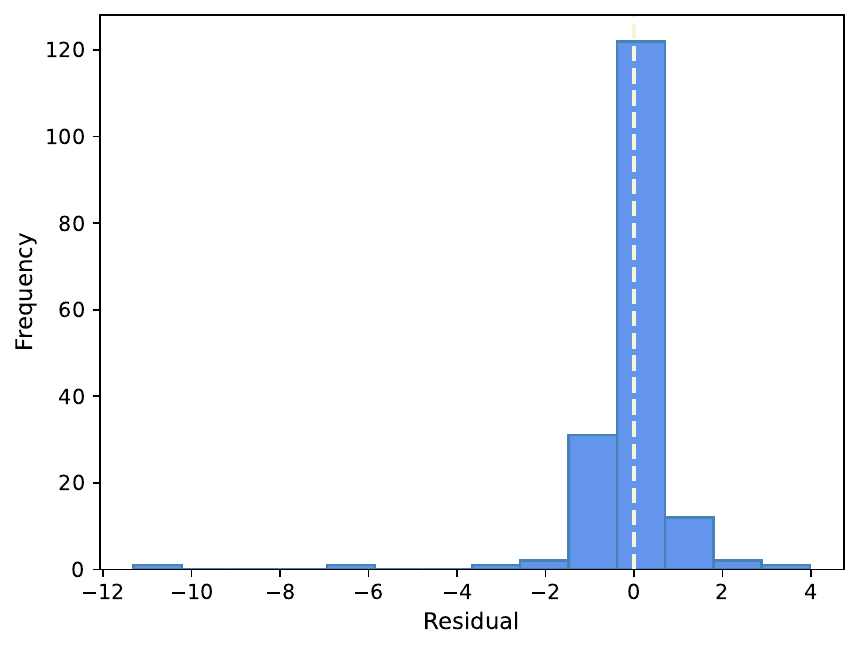}
         \caption{Histogram showing the frequency of overestimated and underestimated predictions.}
         \label{fig:RFEM}
     \end{subfigure}
     \begin{subfigure}{0.4\textwidth}
         \includegraphics[width=\textwidth]{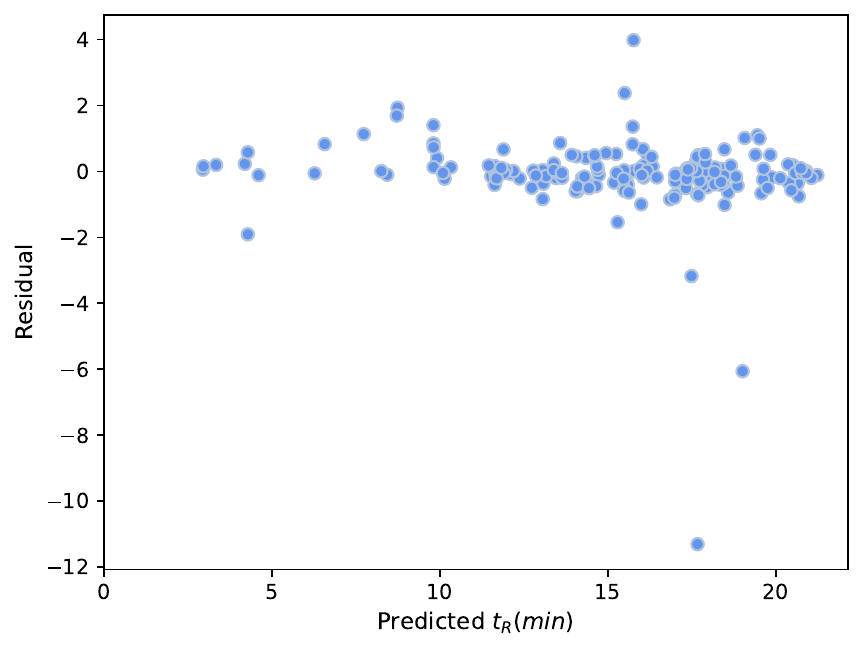}
         \caption{Residual plot showing the difference (min) between the observed and the predicted $t_\mathrm{R}$}
         \label{fig:RFRP}
     \end{subfigure}
        \caption{Error Analysis plots referring to the test results of the optimized RF model on G3 dataset.}
        \label{fig:ResidualPlotRF}
\end{figure}

A similar histogram and residual plot are plotted for the RF results as shown in Figure \ref{fig:ResidualPlotRF}. Consequently, the sequences corresponding to the highest residual values are listed in Table \ref{tbl:RFresiduals}. Except for sequence \#4 in Table \ref{tbl:residuals}, all sequences that were greatly over- or underestimated by the SVR were also difficult to accurately predict by the RF model. However, in the case of RF, the list of sequences with large residuals is longer, ranging between -11.3 and 3.98. All the sequences are fully or partially phosphorothioated sequences, of which 40\% have lost a sulfur atom.

\begin{table}
\setlength{\tabcolsep}{2 pt}
\centering
\caption{ ASO sequences with the highest residual values in the G3 dataset. RF predictions are labeled as over- or underestimated based on the calculated negative or positive values of residuals.}\label{tbl:RFresiduals}
\begin{tabular}{clcccc} 
\hline
\textbf{\#} & \textbf{Sequence} & \multicolumn{1}{c}{\textbf{Observed $t_\mathrm{R}$}} & \multicolumn{1}{c}{\textbf{Predicted $t_\mathrm{R}$}} & \multicolumn{1}{c}{\textbf{Residual}} & \textbf{Over/underestimation}  \\ 
& & (min) & (min) & (min) & \\ 
\hline
1           & G*T*G*G*G*T*G*G*G*T*G*G*G*T*G*G*G*T-P=O   & 6.363                  & 17.672                  & -11.308        & overestimated                 \\
2           & A*T*C*A*G*T*A*T*T*A*A*A*A*T*T*T*T*C*A-P=O & 12.951                 & 19.012                  & -6.061         & overestimated                 \\
3           & C*G*C*G*T*G*T*T*T*T*T*A                   & 14.314                 & 17.490                  & -3.175         & overestimated                 \\
4           & A*C*T*A*T*G-P=O                           & 2.371                  & 4.280                   & -1.908         & overestimated                 \\
5           & G*G*C*G*G*G*C*C*A*G*C*A-P=O               & 13.749                 & 15.291                  & -1.54         & overestimated                 \\
6           & C*G*G*T*GCCTGGCCCCC                       & 17.460                 & 18.477                  & -1.016         & overestimated                 \\
7           & A*A*G*G*C*G*G*T*GCCTGGCCCCC               & 20.091                 & 19.076                  & 1.015          & underestimated                \\
8           & T*TTATCGGCGGG*A*A*C*A-P=O                 & 20.552                 & 19.453                  & 1.098          & underestimated                \\
9           & T*A*T*C*T*C*T*A                           & 8.863                  & 7.732                   & 1.131          & underestimated                \\
10          & A*G*AGCCTGCC*C*G*G*C                      & 17.102                 & 15.742                  & 1.359          & underestimated                \\
11          & T*T*A*G*A*A*T*T*A                         & 11.210                 & 9.808                   & 1.402          & underestimated                \\
12          & T*C*G*A*G*A*C*T*A                         & 10.410                 & 8.718                   & 1.692          & underestimated                \\
13          & T*T*A*T*C*T*C*T*A                         & 10.668                 & 8.736                   & 1.931          & underestimated                \\
14          & G*G*T*G*G*G*T*G*G*G*T*G*G*G*T*G*G*G*T     & 17.878                 & 15.501                  & 2.376          & underestimated                \\
15          & A*T*C*G*G*C*G*G*G*A*A*C*A-P=O             & 19.746                 & 15.765                  & 3.980          & underestimated                \\
\hline
\end{tabular}
\end{table}

\begin{figure}
\centering
    \begin{subfigure}{0.42\textwidth}
         \includegraphics[width=\textwidth]{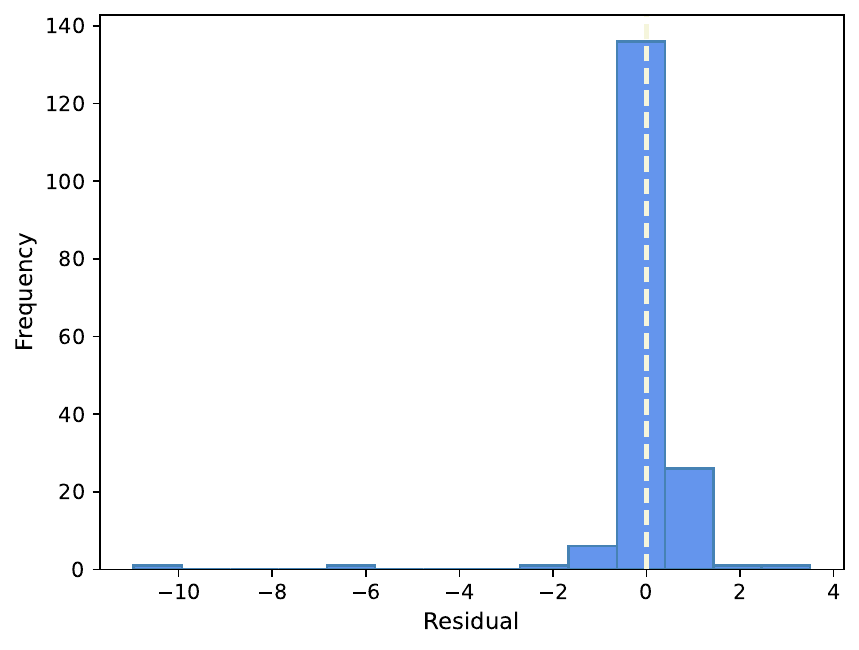}
         \caption{Histogram showing the frequency of overestimated and underestimated predictions.}
         \label{fig:GBEM}
     \end{subfigure}
     \begin{subfigure}{0.4\textwidth}
         \includegraphics[width=\textwidth]{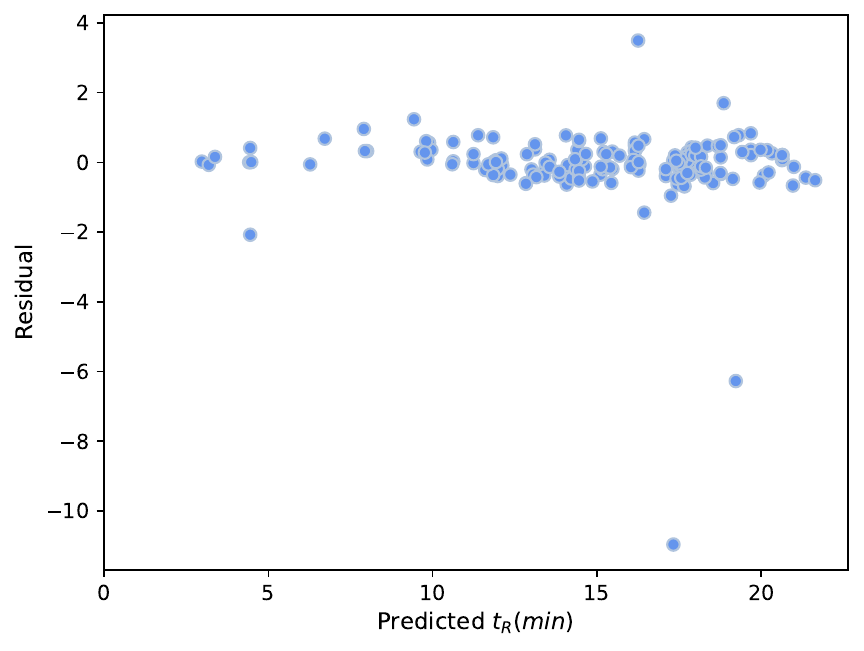}
         \caption{Residual plot showing the difference (min) between the observed and the predicted $t_\mathrm{R}$}.
         \label{fig:GBRP}
     \end{subfigure}
        \caption{Error Analysis plots referring to the test results of the optimized GB model on G3 dataset.}
        \label{fig:ResidualPlotGB}
\end{figure}

In Figure \ref{fig:ResidualPlotGB}, both the histogram and the residual graph are presented for the GB estimator that performs the best. Sequences commonly recognized by SVR as difficult to predict also had high residual values in the GB results, as shown in Table \ref{tbl:GBresiduals}.

\begin{table}
\setlength{\tabcolsep}{2 pt}
\centering
\caption{ ASO sequences with the highest residual values in G3 dataset. GB predictions are labeled as over- or underestimated based on the calculated negative or positive values of residuals.}
\label{tbl:GBresiduals}
\begin{tabular}{clcccc} 
\hline
\textbf{\#} & \textbf{Sequence} & \multicolumn{1}{c}{\textbf{Observed $t_\mathrm{R}$}} & \multicolumn{1}{c}{\textbf{Predicted $t_\mathrm{R}$}} & \multicolumn{1}{c}{\textbf{Residual}} & \textbf{Over/underestimation}  \\ 
& & (min) &(min) & (min) & \\ 
\hline
1           & A*T*C*A*G*T*A*T*T*A*A*A*A*T*T*T*T*C*A-P=O & 12.951                 & 19.223                  & -6.272         & overestimated                 \\
2           & G*T*G*G*G*T*G*G*G*T*G*G*G*T*G*G*G*T-P=O   & 6.363                  & 17.325                  & -10.962        & overestimated                 \\
3           & A*C*T*A*T*G-P=O                           & 2.371                  & 4.447                   & -2.076         & overestimated                 \\
4           & A*G*A*G*C*C*T*G*C*C*C*G*G*C-P=O           & 14.996                 & 16.438                  & -1.442         & overestimated                 \\
5           & T*T*A*T*C*T*C*T*A                         & 10.668                 & 9.432                   & 1.235          & underestimated                \\
6           & T*TTATCGGCGGG*A*A*C*A-P=O                 & 20.552                 & 18.855                  & 1.697          & underestimated                \\
7           & A*T*C*G*G*C*G*G*G*A*A*C*A-P=O             & 19.746                 & 16.251                  & 3.494          & underestimated                \\
\hline
\end{tabular}
\end{table}

\section{Conclusions and Outlook} \label{sec:conclusion}

This paper addresses the challenge of predicting $t_\mathrm{R}$ for ASOs in IPC, a critical step in compound identification and impurity analysis. For this purpose, the performance of four ML models, SVR, GB, RF, and DT, on three large ASO datasets under varying gradient conditions was investigated. Our results show that SVR achieved the highest prediction accuracy and lowest error across datasets, but required significantly more tuning time. GB offered comparable performance, particularly in low-noise datasets, and was at least five times faster. Newly introduced features, such as SuCOUNT and POSITION, improved model performance across the board.

In contrast to earlier studies in literature that were limited to small datasets and single-model evaluations, our comparative analysis offers a broader perspective on multi-model effectiveness for the prediction of $t_\mathrm{R}$ at scale. The results show that the models incorporating multiple features consistently outperformed those relying solely on COUNT for $t_\mathrm{R}$ prediction across all tested models. In G3 dataset, the three ML models consistently struggled with sequences longer than 20 nucleotides, fully phosphorothioated, and lost a sulfur atom.

The presented models offer promising news for the industrial production, separation, and purification of ASOs. First, the good accuracy of $t_\mathrm{R}$ prediction allows efficient sample characterization. Based solely on the ASO sequence data, the predicted $t_\mathrm{R}$ can be used to identify a chromatographic peak. The additional availability of mass spectrometric information confirms and solidifies the assignment. Second, and perhaps more important from a sustainability point of view, the ability to predict ASO separation profiles based on sequence reduces the need for extensive physical experiments. The application of $t_\mathrm{R}$ prediction models may go in lock-step with the development of analytical chemistry methods. The challenges that remain lie in the diversity of ASOs and related compounds; therefore, future research should focus on advanced ML-driven feature engineering, particularly for ASO conjugates and other novel entities emerging in the drug discovery and development field.


Some of the threats to validity encountered during this research are outlined in this section. One such threat concerns model performance, which is inherently dependent on the dataset used, potentially affected by noise introduced during experimental trials. Outliers and noise may correspond to some of the wrongly selected peaks during the data collection phase, which imposes an uncertainty risk on the achieved results. However, it is assumed that the models avoid increasing their complexity so as not to explain the special noise cases. Additionally, the nature of the oligonucleotide data, in particular the relationship between shorter and longer sequences, may influence the performance of the ML models.

\section*{Acknowledgement}
This work was supported by the Swedish Knowledge Foundation via the KKS SYNERGY project “Improved Methods for Process and Quality Controls using Digital Tools—IMPAQCDT” (grant number 20210021). In this project, we are grateful to Gergely Szabados and Patrik Forssén from the Department of Engineering and Chemical Sciences, Karlstad University for their contribution to the experimental work and data preprocessing.


\bibliographystyle{cas-model2-names}
\bibliography{cas-refs}




\end{document}